\documentclass[fleqn,10pt]{wlscirep}
\usepackage[utf8]{inputenc}
\usepackage[T1]{fontenc}

\title{Spin-orbital excitations encoding the magnetic phase transition in the van der Waals antiferromagnet FePS$_{3}$}

\author[1,*,$\dag$]{Yuan Wei}
\author[1,2,*,$\dag$]{Yi Tseng}
\author[3,*]{Hebatalla Elnaggar}
\author[1]{Wenliang Zhang}
\author[1]{Teguh Citra Asmara}
\author[1]{Eugenio Paris}
\author[1,2]{Gabriele Domaine}
\author[1]{Vladimir N. Strocov}
\author[2]{Luc Testa}
\author[2]{Virgile Favre}
\author[4]{Mario Di Luca}
\author[4]{Mitali Banerjee}
\author[5]{Andrew R. Wildes}
\author[6]{Frank M. F. de Groot}
\author[2]{Henrik M. Ronnow}
\author[1,$\dag$]{Thorsten Schmitt}

\affil[1]{Photon Science Division, Paul Scherrer Institut, Villigen PSI, Switzerland}
\affil[2]{Laboratory for Quantum Magnetism, Institute of Physics, École Polytechnique Fédérale de Lausanne, CH-1015 Lausanne, Switzerland}
\affil[3]{Sorbonne Université, CNRS UMR 7590, Institut de Minéralogie, de Physique des Matériaux et de Cosmochimi, 4 Place Jussieu, 75005 Paris, France}
\affil[4]{Laboratory of Quantum Physics, Topology and Correlations, Institute of Physics, École Polytechnique Fédérale de Lausanne, CH-1015 Lausanne, Switzerland}
\affil[5]{Institut Laue-Langevin, 71 Avenue des Martyrs CS 20156, 38042 Grenoble Cedex 9, France}
\affil[6]{Debye Institute for Nanomaterials Science, 3584 CG Utrecht, Netherlands}
\affil[*]{these authors contributed equally to this work}
\affil[$\dag$]{email: yuan.wei@psi.ch; tsengy@mit.edu; thorsten.schmitt@psi.ch}

\begin{abstract}
In the rich phases of van der Waals (vdW) materials featuring intertwined electronic order and collective phenomena, characterizing elementary dynamics that entail the low-energy Hamiltonian and electronic degrees of freedom is of paramount importance. Here we performed resonant inelastic X-ray scattering (RIXS) to elaborate the spin-orbital ground and excited states of the vdW antiferromagnetic insulator FePS$_{3}$, as well as their relation to magnetism. We observed the spectral enhancement of spin-orbital multiplet transitions about $\sim$ 100 and $\sim$ 220 meV, as well as quasielastic response, when entering the zig-zag antiferromagnetic phase, where the spectral changes develop an order-parameter-like evolution with temperature. By comparing with ligand field theory calculations, we discovered the essential role of trigonal lattice distortion and negative metal-ligand charge-transfer to account for these emergent excitations. Such spectral profiles are further examined upon confinement by mechanical exfoliation. We reveal their spectral robustness down to the few atomic layer limit, in accordance with the persistent antiferromagnetic state previously reported in optical measurements. Our study demonstrates the versatile RIXS capability that resolves magneto-crystalline anisotropy and charge-transfer energetics. These provide the crucial insight to understand how the spontaneous magnetic symmetry-breaking stabilizes in the quasi-two-dimensional limit for the vdW magnet FePS$_{3}$.
\end{abstract}

\begin{document}

\flushbottom
\maketitle
\thispagestyle{empty}


\section*{Introduction}
Magnetic van der Waals (vdW) materials have provided exciting new opportunities in the studies of functional exotic magnetic phases of various symmetry-breaking ground states and collective behavior \cite{Bhimanapati2015,Huang2017,Deng2018a,Song2021}. Understanding and controlling the spin state and the magnetic exchange interactions are central to establish the next-generation spin-based opto-electronics based on magnetic vdW materials. Recent studies with both optical X-ray spectroscopy have been utilized to assess the spectral response of the magnetic and electronic configuration down to the few-layer limit \cite{Lee2016,Wang2016,Pelliciari2021a,Lee2022}. These studies resolved the magneto-optical response of the lattice vibrations, and the electronic valence states that give information on spin and orbital configurations, respectively \cite{Lee2016,Wang2016,Lee2022}. However, severe challenges remain for completely understanding the microscopic electronic energy scales in magnetic vdW materials down to the exfoliated few-layer limit. Particularly, many of the exotic spin phases that are of particular interest in realizing spin-based devices occur in the atomically-thin two-dimensional (2D) limit, or in the presence of interlayer translational and angular misfit in vdW heterostructures. Furthermore, dimensional reduction has been shown to host strong electronic correlations. Advanced spectroscopic techniques with sensitivity to the elementary excitations and the electronic order parameter from the bulk to the few-layer limit of vdW materials are of urgent demand.

Here we present a temperature dependent study of the elementary excitations of the vdW antiferromagnetic insulator FePS$_{3}$ using resonant inelastic X-ray scattering (RIXS). FePS$_{3}$ crystallizes in a honeycomb lattice exhibiting a planar zig-zag antiferromagnetic order with out-of-plane moment orientation below 117 K and anisotropic change of the in-plane lattice constants \cite{Lancon2016,Bjarman1983}. Raman studies revealed magneto-optical excitations and phonon modes that signified the magnetic phase transition down to the monolayer limit \cite{Lee2016,Wang2016}. This suggests FePS$_{3}$ as a robust layered nanostructure building block for vdW spintronics. An important outstanding question regards the underlying electronic and spin/orbital structure that enables quasi-2D antiferromagnetism down to the few-layer limit. Such information will guide future research on pathways to tune the low-energy instabilities in FePS$_{3}$, which may provide crucial insight for device applications.

The central question is on the source to stabilize long-range magnetic order in a quasi-two-dimensional lattice environment, which generally requires additional anisotropic interactions governed by the Mermin-Wagner theorem \cite{Mermin1966}. In low-dimensional spin systems, the natural candidates that introduce anisotropy and symmetry-breaking are the magneto-crystalline interactions. Specifically, the zig-zag magnetic order and corresponding local structural nuances are expected to load to spin-lattice coupling \cite{Lancon2016}. Pump-probe optical studies reported the ultrafast control of antiferromagnetic order, which was rationalized by the lowering of orbital symmetry via the trigonal lattice distortion \cite{Ergecen2023}. However, Fe L$_{3}$-edge X-ray absorption spectroscopy (XAS) experiments have remained inconclusive on whether the local C$_{3}$ lattice symmetry-breaking plays a significant role for the emergent magnetic zig-zag orders \cite{Chang2022,Lee2022}. While such local atomic nuances beyond cubic-symmetry octahedral environments was not considered in a recent X-ray study of the isostructural antiferromagnet NiPS$_{3}$ \cite{Kang2020a}, optical pump-probe work suggested the trigonal crystal fields as the leading anisotropic term \cite{Afanasiev2021}. To sum up, the essential elements for stabilizing the antiferromagnetism in FePS$_{3}$ has not been conclusively resolved. 

To resolve these issues, we apply Fe L$_{3}$-edge RIXS to study the low-energy excitations of FePS$_{3}$. RIXS is a spectroscopic technique that probes the dynamics of the elementary excitations in condensed matter \cite{Ament2011c}. The two-step scattering processes of RIXS grant access to a variety of charge-neutral excitations. Transition-metal L-edge RIXS studies have revealed optically-forbidden modes such as single magnon and spin-orbital multiplet excitations \cite{Veenendaal2006a}, and were recently extended to studies investigating few atomic-layer samples \cite{Pelliciari2021,Pelliciari2021a,DiScala2024,Yang2023a}. Particularly, the low-energy spin-orbital excitations have provided valuable information on the low-energy physics of Fe-based correlated materials using RIXS, with strong temperature and polarization dependence due to the presence of the ordered spin and orbital texture \cite{Elnaggar2019b,Elnaggar2020}.

In this work, we report the temperature dependence of the elementary excitations in bulk single crystals of FePS$_{3}$ with Fe L$_{3}$-edge RIXS. Around the Fe L$_{3}$-edge XAS peak maxima (E$_{i}$ = 706.5 and 707.5 eV), we observe two distinguishable modes about $\sim$ 100 and $\sim$ 220 meV that clearly evolve with the magnetic state of the sample. These predominant spectral characteristics clearly suppress upon heating above the Neel temperature, with a crossover-like behavior for the observed excitations. By comparison to the ligand-field calculations with charge-transfer interaction configuration, we assign these excitation modes as two major groups of spin-orbital ${}^1A_{1}$ and ${}^5E$ excited states with lifted degeneracy above the ${}^5A_{1}$ ground state. Importantly, we find that both the trigonal C$_{3}$ lattice distortion and negative metal-ligand charge-transfer are mandatory to reproduce the observed doublet-peak profile with physically reasonable electronic parameters, as well as their temperature developments across the magnetic phase transition. Lastly, spectral comparison of bulk with exfoliated few-layer samples reveal robustness of the electronic structure down to 5 monolayers (MLs), connecting to the reported persisting antiferromagnetism in few-layer samples. Our findings provide a benchmark for RIXS studies in vdW materials hosting magnetic order in general, and shed light to the exotic spin-orbital ground/excited state properties of FePS$_{3}$.

\section*{Results}
\subsection*{Overview of the experimental RIXS results}
\begin{figure}[htbp]
\centering
\includegraphics[width=\linewidth]{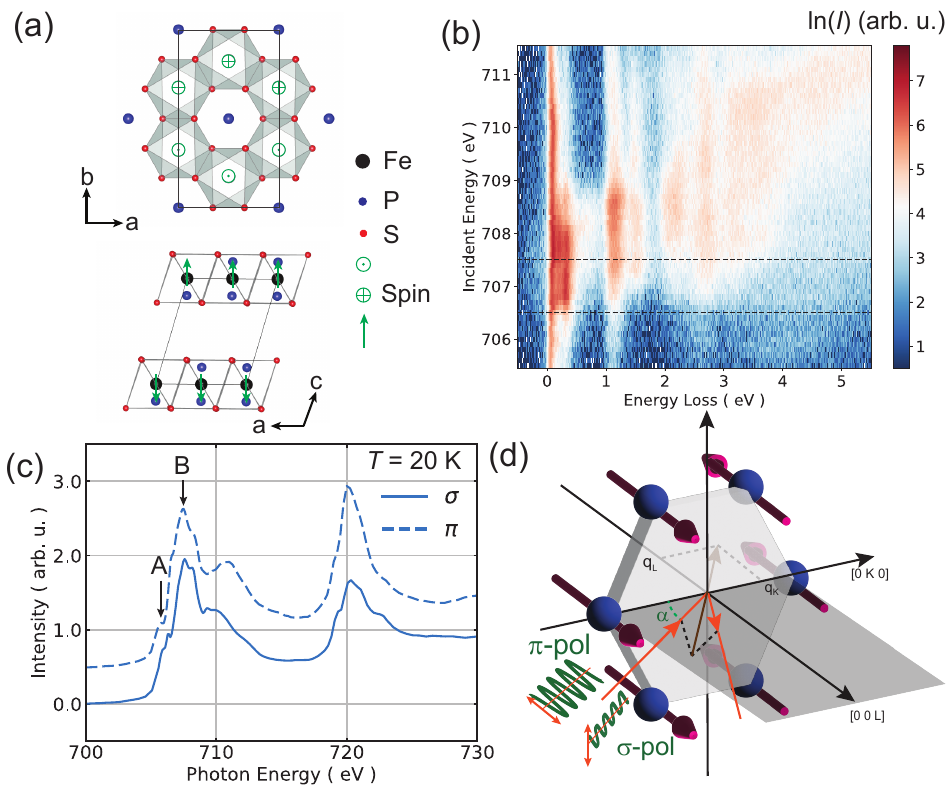}
\caption{\textbf{Crystal structure, magnetic order, XAS and energy-dependent RIXS overview, and scattering geometry.} a, Geometry of the Fe honeycomb plane and interlayer antiferromagnetic structure, where green arrows show the direction of spin. b, RIXS map of FePS$_{3}$ as a function of incident photon energy (left axis) and energy loss (bottom axis). $\pi$ polarization is employed in this measurement. The RIXS intensity is plotted in logarithmic scale. c, Total fluorescence yield (TFY) of FePS$_{3}$ measured at 20 K, with a constant vertical offset to better display the spectra taken with $\sigma$ and $\pi$ polarization, respectively. A = 706.5 eV and B = 707.5 eV mark the two main incident energies of interest as also highlighted in the RIXS map in b (black dashed horizontal lines). d, Experimental geometry of RIXS set up and the local structural environment, see text for details. The TFY and RIXS spectra were measured at the same incident angle $\alpha$ = 12.5$^{\circ}$.}
\label{fig1}
\end{figure}

In FePS$_{3}$, the Fe atoms are arranged as a honeycomb lattice with stacked layers as shown in Fig. \ref{fig1}(a). The magnetic structure has been confirmed to have a collinear antiferromagnetic structure with the moments normal to the ab planes\cite{Lancon2016}, also shown in Fig. \ref{fig1}(a). The Fe L$_{3}$-edge XAS spectra are shown in \ref{fig1}c, which are broadly in agreement with recent works \cite{Lee2022,Chang2022}. The overall spectral appearance is indicative of a d$^{6}$ + d$^{7}\underline{L}$ electronic configuration of the Fe$^{2+}$ ions, where $\underline{L}$ is denoted as the ligand hole state. In Fig. \ref{fig1}b, we present RIXS measurements as a function of incident photon energies E$_{i}$ across the Fe L$_{3}$-edge XAS resonances at base-temperature 20 K (see same measurements above the antiferromagnetic transition in Supplementary Figure 1). We observed a series of localized Raman-like exciations up to 3 eV and fluorescence-like excitations up to about 5 eV energy loss. In a simplified octahedral crystal-field environment, the Fe$^{2+}$ electronic ground state is formed from a ${}^5D$ group (L = 2), with the ${}^5T_{2}$ ground state and first excited state of ${}^5E$ symmetry in the weak-field limit \cite{Grasso1991,Joy1992}. Indeed, the localized peaks about 1 and 1.3 eV loss resembled the spin-allowed ${}^5T_{2g}$ to ${}^5E_{g}$ transitions \cite{Grasso1991,Joy1992}, with a suggested broadening for the 1 eV peak from the dynamical Jahn-Teller effect \cite{Freeman1969}. In the higher-energy regime of 3-5 eV loss, we observe a fluorescence-like spectral response that is resonant about the post-edge broad band in XAS (E$_{i}$ = 710 eV). These are reminiscent of the charge-transfer processes that were previously attributed from the S 3p$_{x}$p$_{y}$ orbitals to the empty Fe 3d shells, as well as the transitions from P-P 3p$_{z}$ to S 3p$_{z}^{*}$ states \cite{Piacentini1982}. At lower-energy loss shown in Fig. \ref{fig1}(c), we observed a peculiar spectral characteristics below 400 meV. These inelastic modes of near-infrared energy range were not covered in previous theoretical or experimental studies. To understand their nature and interplay with magnetism, we performed further temperature dependent RIXS experiments.

\subsection*{Temperature dependence of the experimental RIXS results}
\begin{figure}[htbp]
\centering
\includegraphics[width=\linewidth]{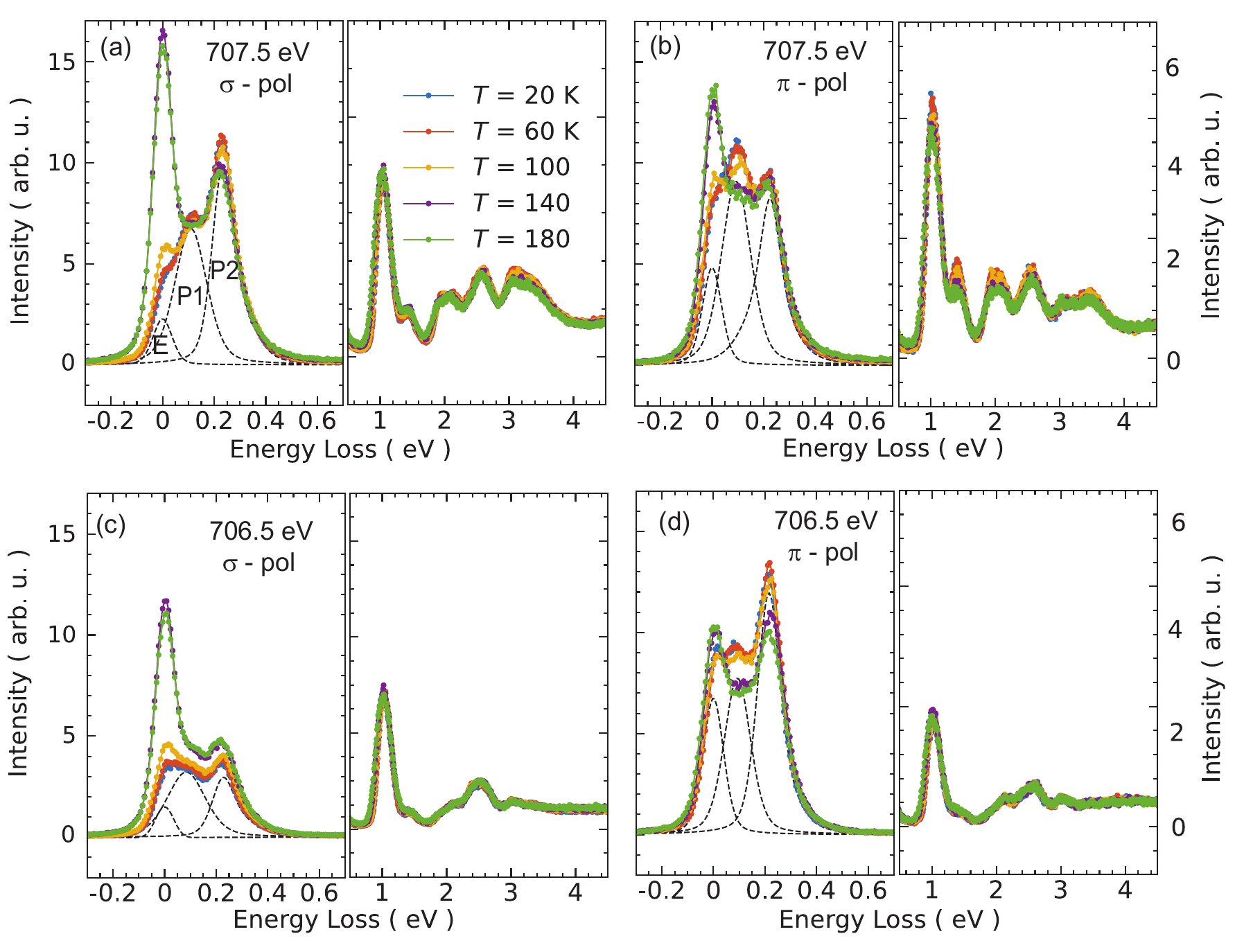}
\caption{\textbf{Temperature-dependent RIXS spectra.} RIXS energy transfer spectra at different temperature with 707.5 and 706.5 eV incident energy in a,b and c,d; spectra recorded for $\sigma$ and $\pi$ incident X-ray polarization are shown in a,b and c,d, respectively. Each graph has panels for the low-energy excitations below 0.6 eV including the elastic peak E and spin-orbital peaks P1/P2 (left) and the high-energy excitations up to 4.5 eV (right). All data was obtained at an experimental geometry with 12.5$^{\circ}$ incidence angle.}
\label{fig2}
\end{figure}

In Fig. \ref{fig2}, we show the temperature evolution of RIXS spectra taken at E$_{i}$ of 706.5 and 707.5 eV with both $\pi$ and $\sigma$ polarization for the incident X-rays. The full temperature series can be found in Supplementary Figure 2. These two resonant energies correspond to the Fe L$_{3}$-edge XAS maximum and a pre-edge shoulder peak (A and B in Fig. \ref{fig1}c), where the predominant low-energy excitations below 500 meV loss are enhanced in spectral intensities (see RIXS spectra taken at other resonances in Supplementary Figure 3). A zoom into this energy regime is highlighted in the insert for the respective incident photon energy and polarization. We observed three distinguishable peaks centering around zero-energy quasi-elastic scattering (peak E), $\sim$ 100 meV and $\sim$ 220 meV loss. Here we designate the $\sim$ 100 and $\sim$ 220 meV excitations as peak 1 and 2, respectively (see fitting assignment in Supplementary Figure 4). With increasing temperature, we found that peak 1 exhibits clear suppression at both E$_{i}$ of 706.5 and 707.5 eV with $\pi$ polarization. This phenomenon becomes more drastic in vicinity to the magnetic phase transition. While the same evolution can be inferred for peak 2 at E$_{i}$ of 706.5 eV with $\pi$ polarization, it becomes nearly temperature-independent for E$_{i}$ = 707.5 eV. On the other hand, both peak 1 and 2 show less temperature developments with $\sigma$ polarization. The elastic peak E, however, undergoes a steep jump right after entering the high-temperature paramagnetic phase after passing 120 K, which is stronger with $\sigma$ polarization. Lastly, excitations above 1 eV energy loss exhibit weak temperature developments in general.

Here our RIXS observations are indicative of spin-orbital multiplet excitations as previously reported in Fe L$_{3}$-edge RIXS studies on magnetite Fe$_{3}$O$_{4}$ \cite{Elnaggar2019b,Elnaggar2020}. The $\sim$220 meV mode is also broadly consistent with the observed spectral profile in a former neutron study \cite{Rule2009}. These findings resemble the fine structure of active $t_{2g}$ orbital energy levels, which are sensitive to exchange fields, spin-orbit coupling and lattice distortions \cite{Wang2018,Benckiser2008,Elnaggar2020,Saha-Dasgupta2019}. The latter naturally connects to the discontinuous decrease (increase) of magnitude in lattice constant \textbf{a} (\textbf{b}) at the magnetic transition \cite{Jernberg1984}, suggesting a possible magneto-restriction mechanism for the observed temperature developments of the low-energy excitations peak 1 and 2. Such strong spin-lattice coupling is evidenced by the previous high-field magnetization experiments and anisotropic exchange model calculations \cite{Wildes2020,Wildes2020a}. Meanwhile, the corresponding incidence angle dependence with light polarization follows previous studies on multiplet transitions of a 3d$^6$ configuration (see Supplementary Figure 5). On the other hand, the differences in the details of the temperature dependence (independence) for peak 2 taken at E$_{i}$ = 706.5 (707.5) eV shown in Fig. \ref{fig2}(a)-(b) might relate to the different initial and intermediate states involved in the RIXS processes \cite{Ament2011c}. As for the elastic peak E enhancement upon heating, this phenomenon reminds us of the increasing weight for the spin-singlet ${}^1A_{1}$ state in other reports \cite{Miao2019a,Marino2023}. It was rationalized that the high-temperature paramagnetic phase could favor such raised symmetry for s-wave like spin and orbital states \cite{Miao2019a,Marino2023}. This scenario, however, would require further investigations since the RIXS elastic signal typically involves multiple contributions, e.g. resolution-limited excitations, acoustic phonons, diffuse scattering, etc \cite{Ament2011c} (see Supplementary Section 4.1).

\subsection*{Calculated electronic and spin-orbital states in the RIXS spectra}

\begin{figure}[htbp]
\centering
\includegraphics[width=\linewidth]{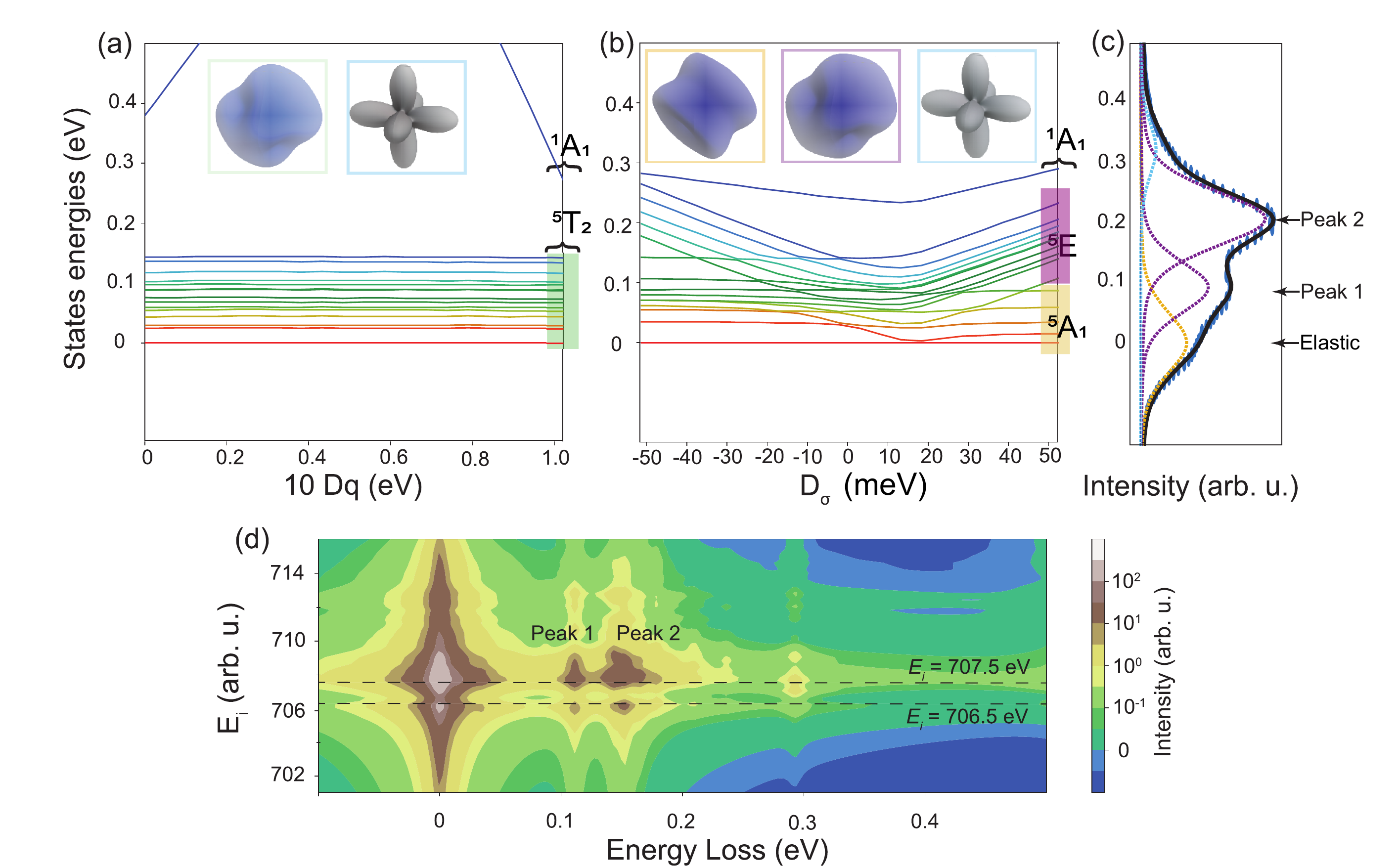}
\caption{\textbf{Multiplet ligand field theory calculations} Energy level diagram of multiplet states for Fe$^{2+}$ ions in FePS$_3$ together with the RIXS spectral response determining the distortion parameters. a, The effect of $10D_q$ parameter representing the octahedral crystal field varied from zero to 1 eV. b, The effect of $D_{\sigma}$ parameter representing the magnitude of trigonal distortion. c, An experimental low energy RIXS spectrum for an incident energy of 707.5 eV and $\sigma$ polarization is plotted aligned to the energy level splittings predicted by our calculations. The multiplet parameters were optimized to quantitatively describe the experimental data (blue corsses) afrer experimental broadening. d, Full Fe 2p3d RIXS map calculations around the Fe L$_3$-edge for FePS$_3$. The calculated intensities are averaged over $\pi$ and $\sigma$ polarization for the outgoing X-rays.}
\label{fig3}
\end{figure}

To understand our RIXS inelastic response, we performed charge-transfer multiplet (CTM) theory calculations to gain an accurate understanding of the electronic structure of FePS$_3$ and pinpoint to the interactions leading to peak 1 and 2. The ground state of an Fe$^{2+}$ ion (3d$^{6}$) in cubic symmetry is $^5T_2$ which is composed of 15 micro-states that are split due to magnetic exchange interaction. The energy level diagram of Fe$^{2+}$ taking into consideration three charge transfer configuration (i.e. d$^{6}$ + d$^{7}\underline{L}$ + d$^{8}\underline{L}^2$ with negative charge transfer) as a function of the octahedral crystal field parameter, $10D_q$ is shown in Fig. \ref{fig3}a). It is expected that the octahedral crystal field is $\sim$1 eV for TMPS$_3$ compounds (TM = Mn, Fe, Co, Ni) \cite{Kang2020a}. This leads to a situation where the 15 states belonging to the $^5T_2$ spans the energy range 0 - 150 meV followed by the $^1A_1$ state at $\sim$300 meV. Density plots of representative wave-functions belonging to the $^5T_2$ and $^1A_1$ multiplets are shown at the top of Figure \ref{fig3}a and \ref{fig3}b and are color coded respectively. The main effect of the $10D_q$ parameter can be observed for the $^1A_1$ state, where increasing $10D_q$ initially increases the energy of $^1A_1$ and then decreases it eventually towards the ground state for large $10D_q$ values. 

We concluded that an extra interaction must be at play and that the Fe ions must be distorted. This is a new aspect that was not observed for NiPS$_3$ \cite{Kang2020a} and agrees with the very recent results of X-ray photoemission electron microscopy on FePS$_3$ \cite{Lee2022}. This trigonal distortion splits the $^5T_2$ multiplet to $^5A_1$ (5 states) and $^5E$ (10 states) as shown in Fig. \ref{fig3}b. We found that a distortion of $\sigma = 50$ meV explains our experimental data well where the (quasi) elastic peak is comprised of spin excitations belonging to the $^5A_1$ manifold while peaks 1 and 2 are comprised of spin-orbital excitations belonging to the $^5E$ manifold (Fig. \ref{fig3}b,c). The full theoretical RIXS map is shown in Fig. \ref{fig3}d and captures our experimental results very well.

Additionally, we conclude that both the presence of negative charge-transfer and the trigonal lattice environments are essential to reproduce our experimental observation. This is examined by systematically adjusting the values of exchange interaction field, crystal-field parameter $10D_q$, and the charge-transfer $\delta$ that corroborate with the existing literature of FePS$_{3}$. While a sizable metal-ligand covalency was speculated in FePS$_{3}$ in a recent XAS study \cite{Chang2022}, our RIXS results reveal the importance of invoking the negative metal-ligand charge-transfer for explaining the excitation dynamics.

\begin{figure}[htbp]
\centering
\includegraphics[width=\linewidth]{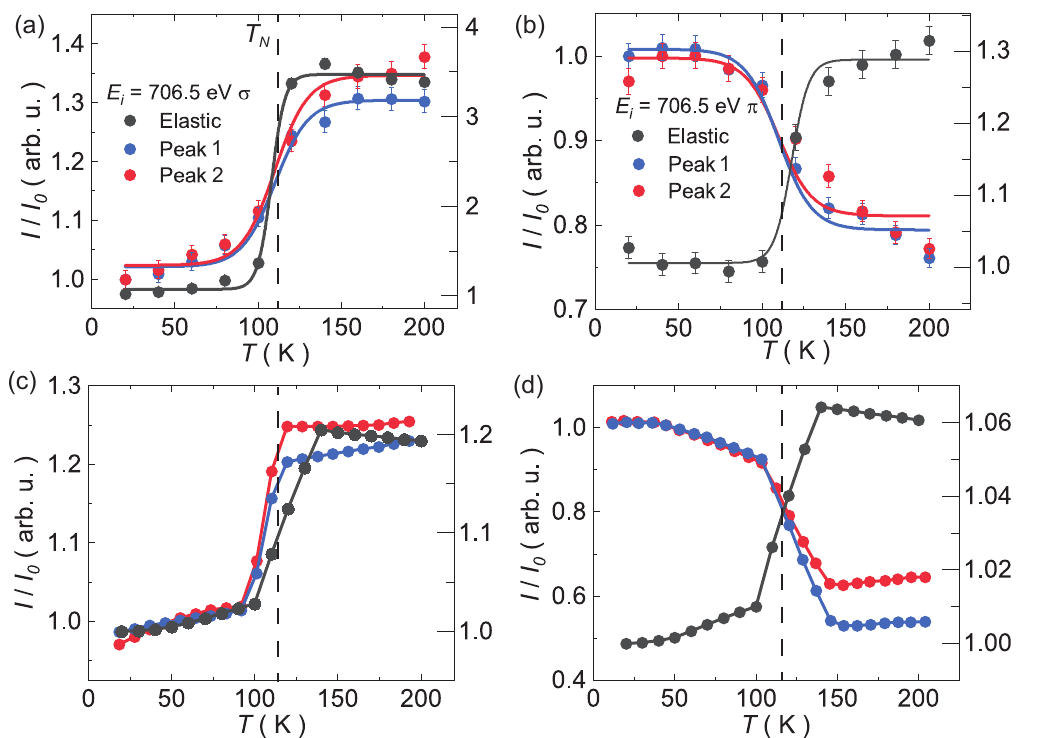}
\caption{\textbf{Comparison between experimental and theoretical RIXS response vs. temperature.} Temperature dependence of the RIXS intensity of the spin-orbital excitations across the magnetic phase transition temperature. All intensity values are represented relative to the value at 20 K for comparative analysis. Experimental RIXS spectra are taken at $E_{i}$ = 706.5 eV. a-b, Experimental integrated intensities of peak 1, peak 2 and the elastic peak. The intensity scale of the spin-orbital peaks and the elastic line are at the left and the right axis, respectively. c-d, The integrated intensity amplitudes of peak 1 and 2 as well as E from theoretical calculations. The antiferromagnetic ordering temperature $\sim$120 K is indicated by a black dashed line in all panels. The errors are standard deviations of fitting results.}
\label{fig4}
\end{figure}

We then compare the temperature evolution of the experimental RIXS response and the calculated spin-orbital excitations peak 1 and 2,  as well as the elastic scattering, across the magnetic phase transition in Fig. \ref{fig4}. In these calculations, we assume the molecular exchange field stays constant in the antiferromagnetic phase and is effectively zero in the paramagnetic phase. The Boltzmann occupation in the ground state configuration is taken into consideration. With these assumptions, our theoretical model captures the main essence of our experimental observations: (1) The intensities of the spin-orbital excitations and elastic line are strongly modified at the magnetic transition temperature, exhibiting a two-state like crossover. (2) The order of magnitude in intensity changes and effects of light polarization are captured. The remaining differences, including some intensity background sources that gradually change with temperature without a steep trend at the magnetic phase transition, may be attributed to additional factors like Boltzmann population broadening via phonon coupling or short-range spin fluctuations. These discrepancies may arise due to the simplifications in our model. Notably, our calculations assume that all magnetic interactions are completely missing above the transition temperature, which is an approximation rationalized by the melting of the low-energy spin-wave excitations from a previous neutron study \cite{Lancon2016}. Nevertheless, neutron studies have shown a Lorentzian-like quasi-elastic weight extending to $\sim$40 meV and persisting above the antiferromagnetic transition, suggesting the possible non-negligible spin fluctuations in the paramagnetic phase \cite{Wildes2012}. The other extreme condition is to keep the exchange interactions of the same magnitude above and below the transition when calculating the RIXS intensity as a function of temperature (see Supplementary Section 4.2 and Supplementary Figure 7). Nevertheless, our experimental RIXS response can be reasonably captured by CTM theory with a two-state crossover across the magnetic phase transition, signifying an order-parameter-like development.

On the other hand, the elastic intensity increase across the antiferromagnetic transition is also qualitatively captured in the calculations. A possible explanation for the small differences in experiments compared to the simulations may involve the spin-singlet multiplet states around zero-energy. This was observed in low-energy multiplet excitations near zero-energy loss, where the spin- and orbital-singlet components got strengthened in weight in the high-temperature paramagnetic phase \cite{Miao2019a,Marino2023}. Nevertheless, our extended CTM calculations predict that additional spectral modes to appear in the spin-orbital multiplet spectrum if this picture is at play (see Supplementary Figure 6), which motivates future more advanced theory developments to completely describe the low-energy response. 

\subsection*{Exfoliated thin flake RIXS response}
\begin{figure}[htbp]
\centering
\includegraphics[width=\linewidth]{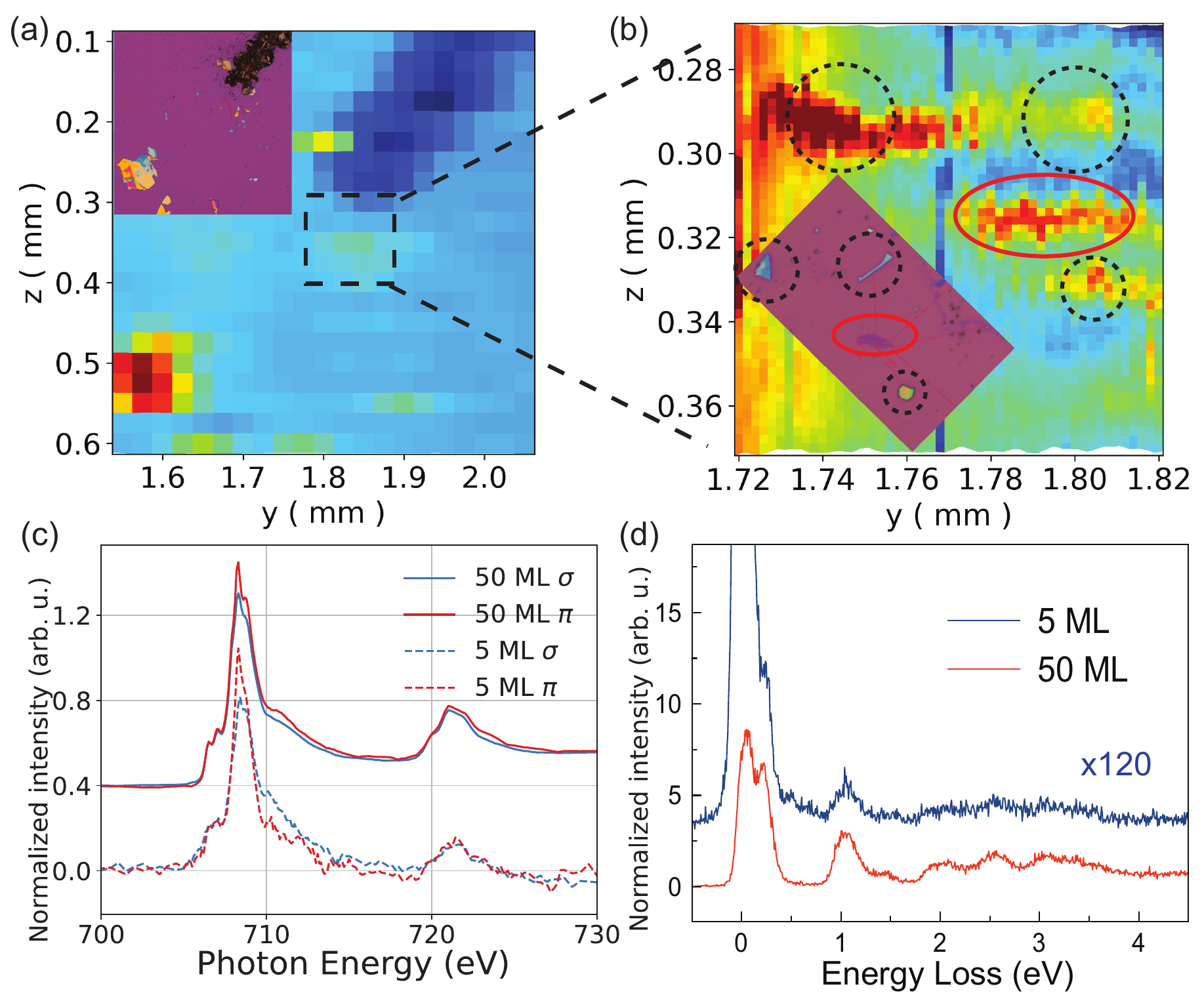}
\caption{\textbf{Exfoliated thin flake XAS and RIXS response.} a, The optical microscopy image (insert top left) and spatial map of total fluorescence yield at Fe L$_{3}$-edge absorption maximum for position registry of the mechanically exfoliated FePS$_{3}$ flake samples with 5 ML thickness on a SiO$_{2}$/Si substrate. b, Zoom-in region around the measured flakes with the corresponding optical image and XAS spatial map. The 5 ML sample and the surrounding thicker bulk-like flakes, including the thickest near the top left corner (50 ML), are highlighted by the red solid-line ellipse and black dotted-line circles, respectively. c-d, XAS and RIXS data recorded on exfoliated 50 ML and 5 ML samples at 20 K. RIXS data for 5 ML sample is scaled up in spectral intensity for clarity in comparison.}
\label{fig5}
\end{figure}

Lastly, we exploit the response of these spin-orbital excitations towards the 2D limit by employing RIXS measurements in mechanically exfoliated samples. In Fig. \ref{fig5}a-b, optical microscopy images and XAS spatial mapping of the exfoliated samples are shown and cross-compared for position registry. The sample flake preparation and thickness characterization can be found in Supplementary Section 5 and Supplementary Figure 8. The corresponding XAS and RIXS spectra are shown in Fig. \ref{fig5}c-d. We observe that the overall XAS spectra and RIXS profiles extending to higher-energy excitations persist in the exfoliated flakes of 50 atomic layers (50 ML) and atomically thin 5 MLs thickness. This shows that the multiplet excitation response, and thus the electronic structure, of the bulk phase remains intact in the exfoliated thin flakes. The low-energy peak 1 of the bulk phase ($\sim$100 meV) is more difficult to assess in few-layer samples of 5 MLs due to the substrate diffuse (elastic) scattering and the correspondingly lower signal level. Nevertheless, the persistence of the multiplet excitation response reflects onto to the robust antiferromagnetism in FePS$_{3}$ down to the few-layer limit as in previous optical measurements \cite{Wang2016}.

\section*{Discussion}

We demonstrate the versatile RIXS sensitivity to low-energy excitations in vdW antiferromagnet FePS$_{3}$ from spin-orbital multiplet transitions that grant access to the microscopic electronic interactions. The temperature- and polarization-resolved RIXS measurements shows an evolution of the low-energy dynamics at $\sim$100 meV and $\sim$220 meV that correlates with the development of the magnetic order parameter and the underlying electronic wavefunction. With the understanding of the dynamics of the elementary excitations registering the magnetic phase transition, our study allows the assessment of the crucial role of trigonal lattice distortions and the negative metal-ligand charge transfer. With these findings, we capture the main anisotropic interactions that stabilize the long-range antiferromagnetism in the quasi-2D limit, which we find to persist in our RIXS and XAS measurements comparing the bulk crystals and exfoliated 5 ML flake samples. Our investigation resolves the ground and excited state wavefunctions with their detailed spin/orbital texture and corresponding symmetries. The control these ground state properties is a prerequisite for material engineering utilizing spin-lattice interactions and charge-transfer energetics, e.g. piezo-control electronics with strain tuning, spin-flip photoluminescence sensors in OLEDs and quantum sensing \cite{Kitzmann2022}. Our work highlights RIXS as an ideal generalized approach for studying functional 2D materials, featuring sensitivities to all degrees of freedom and flexible parameter control.

\section*{Methods}

\subsection*{Resonant inelastic X-ray scattering experiment}
Fe $L_{3}$-edge resonant inelastic X-ray scattering (RIXS) and X-ray absorption spectroscopy (XAS) experiments were performed at the Advanced Resonant Spectroscopies (ADRESS) beamline of the Swiss Light Source at the Paul Scherrer Institut \cite{Strocov2010a,Strocov2011a,Ghiringhelli2006a}. The total energy resolution of the RIXS experiment was about 85 meV at the Fe $L_{3}$-edge ($\approx$ 707 eV). The RIXS spectrometer was fixed at a scattering angle 2$\theta$ = 130$^{\circ}$. An experimental geometry was used that was fixing the scattering plane in the bc plane with the in-plane momentum transfer along the crystallographic [010] direction, with an incidence angle of 12.5$^{\circ}$. RIXS spectra were acquired with 15 (1) minutes for the temperature (incident energy) dependent measurements, respectively, and normalized to the incoming beamline flux. XAS spectra were recorded in total fluorescence yield (TFY) mode. Both $\pi$ and $\sigma$ polarization was employed for the incident X-rays. Unless specified, all RIXS and XAS measurements were performed at the base temperature at 20 K. Details on the synthesis of single crystal FePS$_{3}$ samples can be found in ref. \cite{Lancon2016}.

\subsection*{Multiplet ligand field theory calculations}
To understand the character of the low-energy excitations peak 1 and 2, we employ exact diagonalization calculations within charge-transfer multiplet (CTM) theory as implemented in Quanty \cite{Haverkort2014,Haverkort2012,Lu2014}. The model reduces to a multielectronic calculation of a single FeS$_6$ cluster under $O_h$ symmetry, accounting for the Fe-$3d$ orbitals and the corresponding symmetrized molecular orbitals from Fe 3d states with hybridization to S 3p states \cite{Haverkort2014,DeGroot,VanderLaan1986}. Additional symmetry-breaking by trigonal fields of symmetry breaking is taken into account for quantification. Configuration interaction calculations taking into account (i) the intra-atomic Coulomb interaction, (ii) the crystal field, (iii) charge transfer, (iv) the mean field exchange interactions, and (v) spin-orbit interaction were performed using the quantum many-body program Quanty \cite{Haverkort2014,Haverkort2012,Lu2014}. The $d-d$ and $p-d$ multipole part of the Coulomb interaction was scaled to 80\% of the Hartree-Fock values of the Slater integral. Three charge transfer configurations were taken into account in the calculations ($i.e.$, $3d^n$ + $3d^{n+1}\underline{L}$ + $3d^{n+2}\underline{L}^2$).
The following parameters were used for the multiplet calculations: 
\begin{table}[hb]
    \centering
    \begin{tabular}{|c|c|c|c|c|c|c|c|c|c|} 
            \hline
          (eV)     & $10 D_q$ & $D_{\sigma}$ & $\delta$ & $V_{a1},V_{b1}$& $V_{e},V_{b2}$ & $U_{pd}$ & $U_{dd}$ & $J_{exch}$ & $SOC_{3d}$\\  \hline \hline
      Initial      &  1       &     0.05     & 0.98     & 1.5  & 0.7 & - & 5& 0.012& 0.052\\      
      Intermediate &  1       &     0.05     & -0.55    & 1.5  & 0.7 & 6 & 5& 0.012& 0.0665\\  \hline
    \end{tabular}
    \caption{Input parameter for multiplet calculations of Fe in FePS$_3$. Here the trigonal crystal field parameters are given by $10 D_q$ and $D_{\sigma}$. The charge transfer energy and hybridization are given by and $\delta$ and $V$. The onsite energies are given by $U_{pd}$ and $U_{dd}.$ The mean field exchange field is $J_{exch}$ and the spin-orbit coupling constant is referred to as $SOC_{3d}$. }
    \label{tab:CTMParameters}
\end{table}

\section*{Data availability}
All the data used in the present manuscript and supplementary information are available on request. 

\section*{Code availability}
All the accession codes used in the present manuscript and supplementary information are available on request.

\section*{Acknowledgements}
We thank Jin Jiang and Zekang Zhou for their assistance in the fabrication and characterization of the exfoliation samples. The experiments have been performed at the ADRESS beamline of the Swiss Light Source at the Paul Scherrer Institut (PSI). We acknowledge financial support by the Swiss National Science Foundation through project no. 207904, 178867 and 160765, as well as from the European Union’s Horizon 2020 research and innovation programme under the Marie Sk$\l{}$odowska-Curie grant agreement no. 701647 and 884104 (PSI-FELLOW-III-3i), and Dutch Research Council Rubicon Fellowship (Project No. 019.201EN.010). M.B. acknowledges the support of SNSF Eccellenza grant No. PCEGP2$\texttt{\textbackslash} \textunderscore$194528, and support from the QuantERA II Programme that has received funding from the European Union’s Horizon 2020 research and innovation program under Grant Agreement No 101017733.

\section*{Competing interests}
The authors declare no competing interests.

\section*{Author contributions}
Y.T., H.M.R. and T.S. designed the experiment; Y.W., Y.T., W.Z., T.C.A., E.P., G.D., V.N.S. and T.S. performed the experiment; H.E. and F.M.F.d.G. performed the theoretical calculations; A.W. prepared and characterized the single crystal samples; L.T. and V.F. performed Laue diffraction for aligning the single crystal samples; M.D.L. and M.B. assisted in the preparation of exfoliated thin flake samples; Y.W. performed data analysis in discussion with Y.T., H.E., F.M.F.d.G. and T.S.; T.S. was responsible for project management; Y.W., Y.T., H.E., F.M.F.d.G, and T.S. wrote the paper together with input from all other authors.

\bibliography{FePS3}

\newpage

\end{document}


\maketitle
\bibliographystyle{naturemag}


\section{Additional incident-energy dependent RIXS data}
Here we report additional incident-energy dependent RIXS data. These results are performed in same way as the measurements shown in main text as RIXS energy map (Fig. 1b) and temperature-dependent RIXS linecuts (Fig. 2), respectively. The experimental configuration is fixed at a scattering angle 2$\theta$ = 130$^{\circ}$ for the experimental geometry, with the same bc scattering plane and the in-plane momentum transfer along the crystallographic [010] direction in 12.5$^{\circ}$ grazing incidence. In Supplementary Figure 1, we present the incident-energy dependence of RIXS spectra taken at 200 K, which is well above the antiferroamgnetic ordering temperature $\sim$120 K. Except for the uprising intensity level of the elastic line centering at zero-energy loss, we observe that all the inelastic RIXS modes exhibit an overall weight suppression upon heating.

On top of this, we further explore the temperature developments at other selected incident-energies across the Fe L$_{3}$ X-ray absorption spectroscopy (XAS) profiles. Here we focus on the excitation profiles taken at distinct resonances away from the pre-edge and maximum of Fe L$_{3}$ XAS spectrum as shown in the main text (E$_{i}$ = 706.5 and 707.5 eV). A full temperature series for RIXS spectra taken around the Fe L$_{3}$-edge resonances is shown here for comparison in Supplementary Figure 2. For comparison, the RIXS data taken away from these resonances are summarized in Supplementary Figure 3, where we compare the temperature evolution of RIXS spectra at the major post-edge region at E$_{i}$ = 711 eV. The low-energy multiplet weight below 400 meV, specifically peak 1 $\sim$100 meV and peak 2 $\sim$220 meV that we assigned as spin-orbital excitations governed by trigonal lattice distortions and negative charge-transfer interactions (see main text Figure 2 and Supplementary Figure 1), are much suppressed in intensity and less defined in spectral structure compared the RIXS data taken at E$_{i}$ = 707.5 and 706.5 eV. Meanwhile, the relative weight of higher-energy fluorescence-like weight above 3-4 eV loss are also enhanced compared to the E$_{i}$ = 707.5 and 706.5 eV resonances. Both spectral components below 400 meV and above 3 eV exhibit weaker temperature evolution compared to E$_{i}$ = 707.5 and 706.5 eV as well. This could be rationalized as the enhanced contribution from ligand charge-transfer channels that are enabled in this XAS energy region. This is due to the initial and final state wavefunction configurations at this incident-energy resonance may present different orbital symmetries and hybridization to neighboring ligands, which likely account for the overall reduced sensitivity to spin and lattice nuances across the magnetic phase transition. This again highlights our main text data around the Fe L$_{3}$-edge XAS maximum that shows direct coupling to the fine structure of multiplet dd transitions, such that the involved lattice distortions and spin-orbit interactions give rise to a strong fingerprint to the underlying magnetism.

\begin{figure}[!htbp]
\centering
\includegraphics[width=\linewidth]{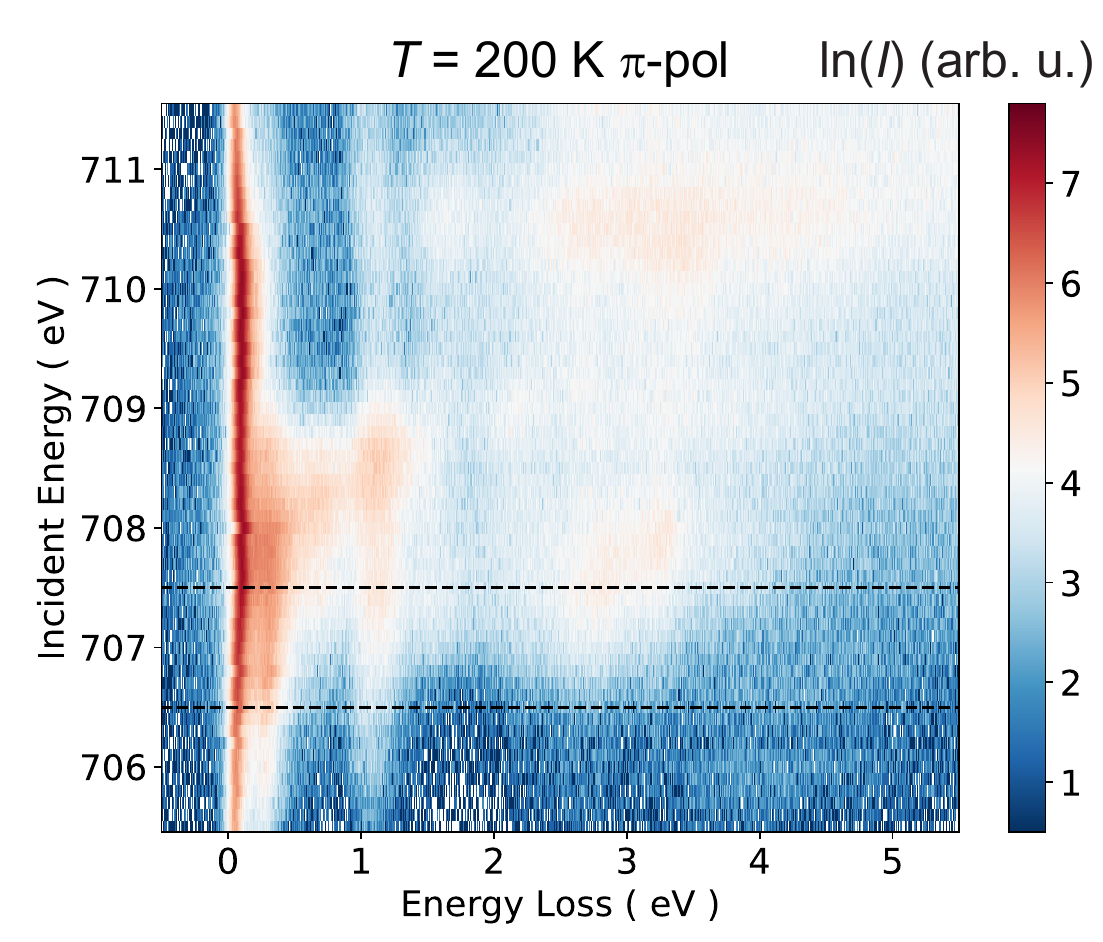}
\caption{Incident-energy dependent RIXS map taken at 200 K with $\pi$ polarization. The RIXS intensity is plotted in logarithmic scale.}
\label{figS4}
\end{figure}


\begin{figure}[!htbp]
\centering
\includegraphics[width=\linewidth]{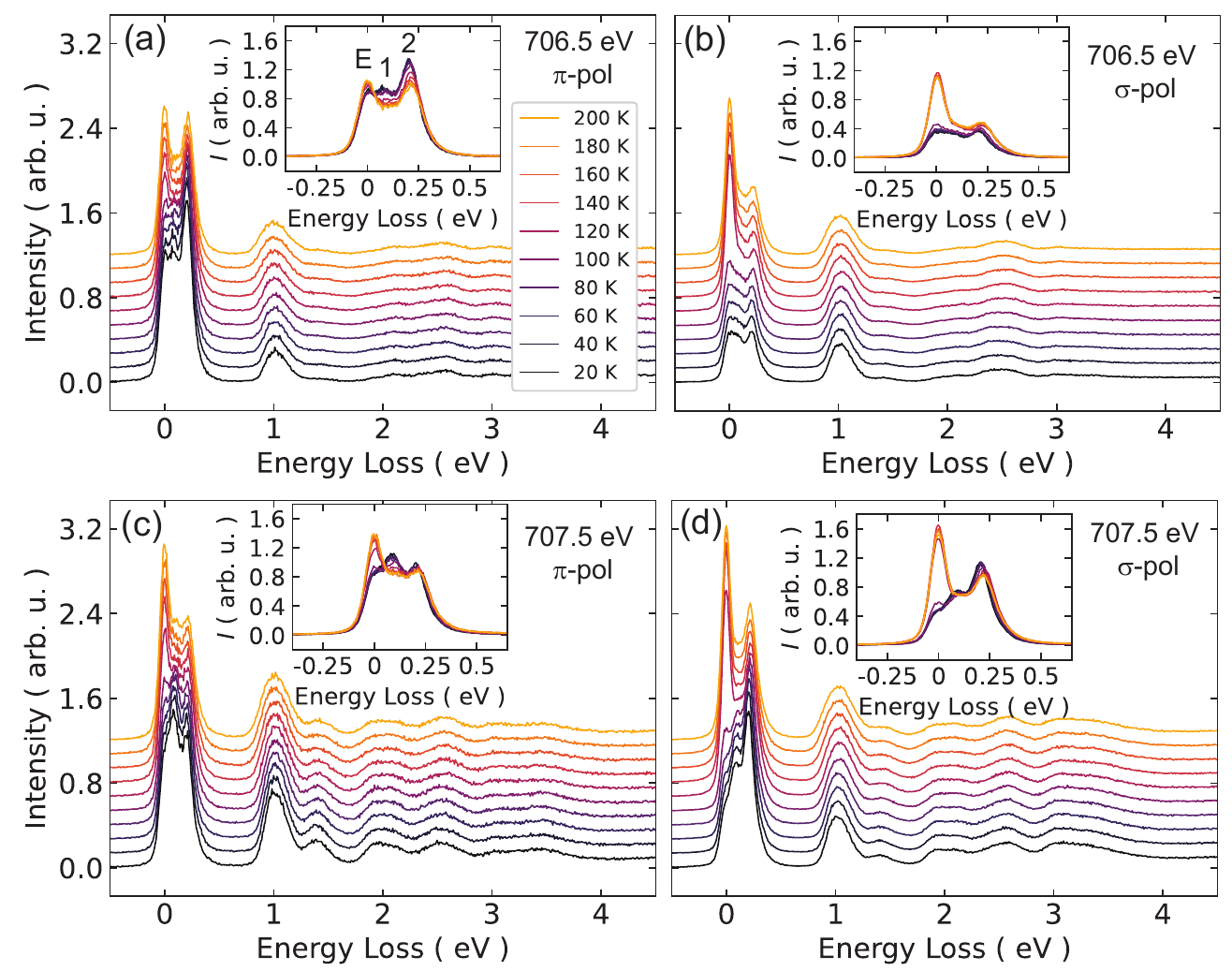}
\caption{Full temperature-dependent RIXS taken around the Fe L$_{3}$-edge XAS maxima.}
\label{fig2}
\end{figure}

\begin{figure}[!htbp]
\centering
\includegraphics[width=\linewidth]{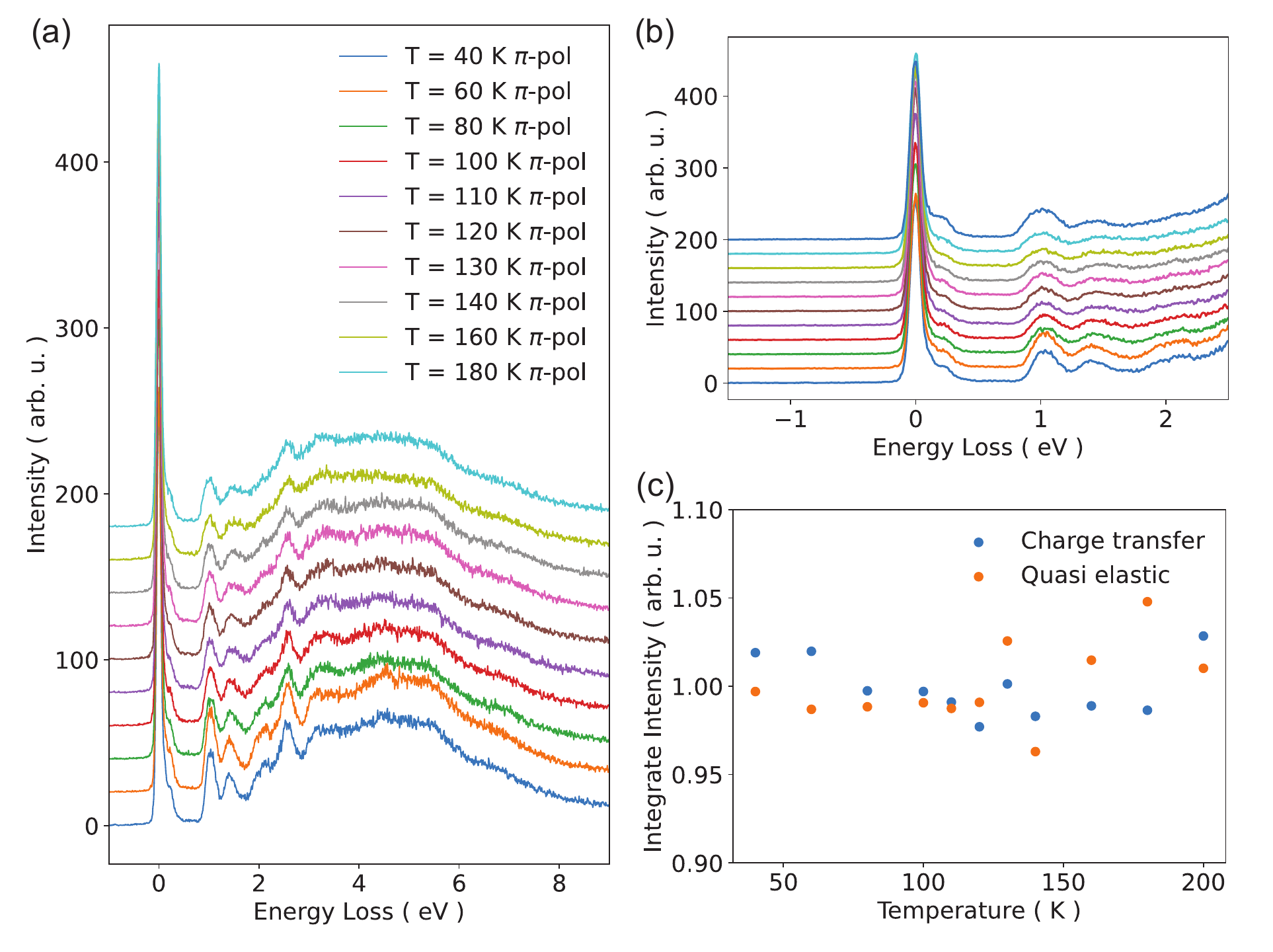}
\caption{RIXS data at $E_{i}$ = 711 eV. (a) A waterfall plot to present a comprehensive overview of the RIXS spectra at different temperatures. (b) Zoom in on the low-energy regime. (c) Integrate the intensity of higher-energy dd and charge-transfer modes, as well as the low-energy excitations (covering peak 1 $\sim$100 eV and peak 2 $\sim$220 eV), respectively. The blue scatters represent the range of 3 to 6 eV, and orange scatters represent the range of 0.1 to 0.4 eV. Intensity is normalized by an average factor for a clearer visualization of overall intensity fluctuations.}
\label{figS7}
\end{figure}


\section{Spectral fitting for RIXS data}
To quantify the spectral weights of the low-energy excitations of interest below 500 meV loss, we applied a fitting model with pseudo-Voigt functions (\ref{eq1}) to describe the spectral profile.
\begin{equation}\label{eq1}
f(A, \mu, \sigma, \alpha) =  \frac{(1 - \alpha)A}{\sigma\sqrt{2\pi}} e^{-[(x-\mu)^2/2\sigma^{2}]} + \frac{\alpha A}{\pi} [\frac{2\sqrt{2} \sigma}{(x-\mu)^{2}+(2\sqrt{2} \sigma)^{2}}]
\end{equation}
in which A is a fitting coefficient. To fit the elastic line, a Lorentzian contribution $\alpha = 0.46 $ is employed for the relative weight between Gaussian and Lorentzian contribution in the pseudo-Voigt function. The full widths at half maximum ($2 \sigma$) of elastic line at zero-energy loss, as well as the spin-orbital multiplets peak 1 ($\sim$100 meV) and peak 2 ($\sim$220 meV), are about 70, 120, and 110 meV, respectively. Finite residual weight is inferred at the asymmetric higher-energy tail of peak 2, which is compensated by a resolution-limited component. The fitting examples at different temperature are shown in Supplementary Figure 4. 

\begin{figure}[!htbp]
\centering
\includegraphics[width=\linewidth]{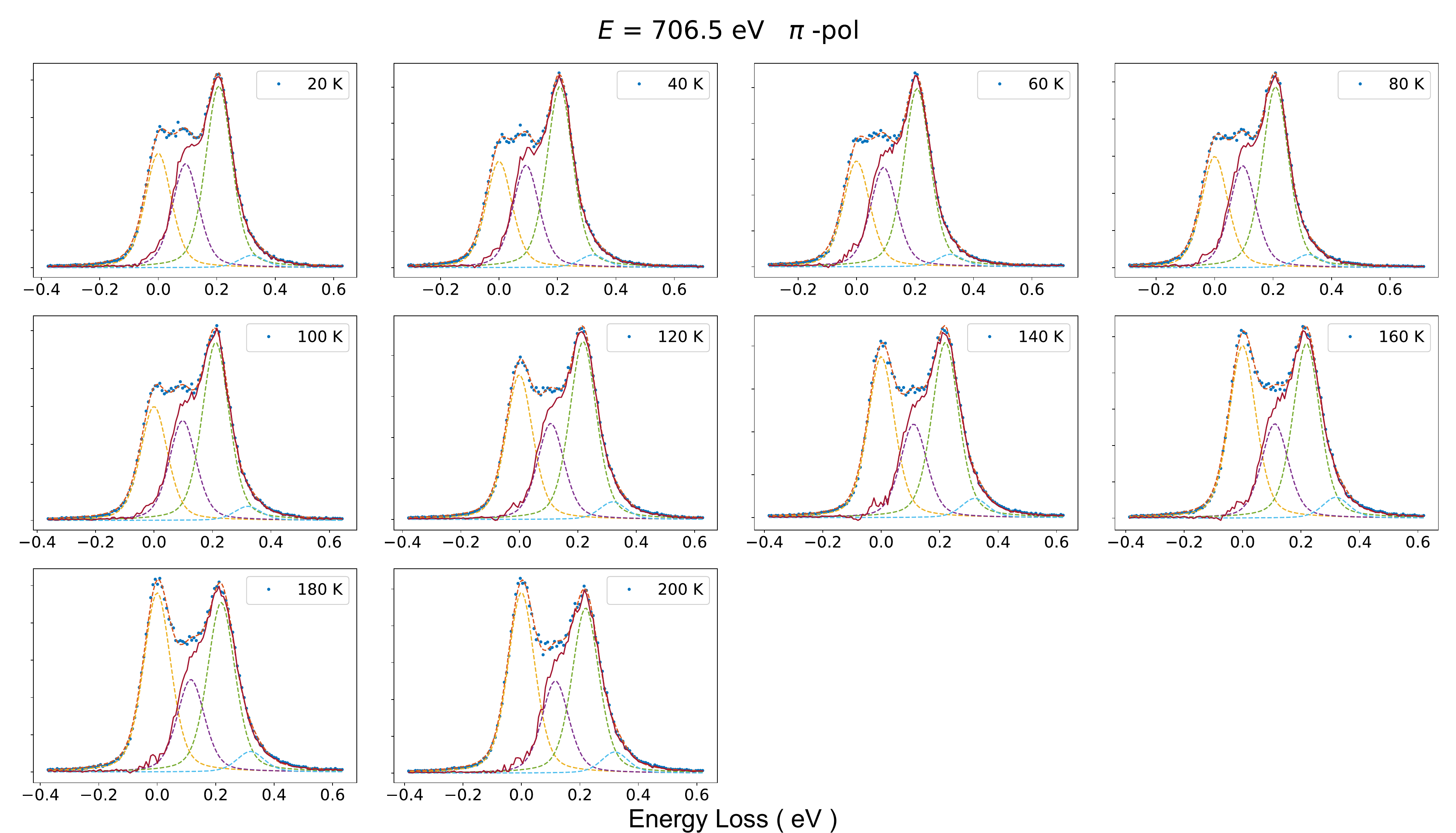}
\caption{Exemplary fitting scheme of RIXS spectra as a function of temperature. The raw data (blue filled markers), as well as spectral components of elastic line (orange dotted line), peak 1 $\sim$100 meV (purple dotted line), peak 2 $\sim$220 meV (green dotted line), and remaining higher-energy tail weight of peak 2 (cyan dotted line) are shown respectively. A elastic-subtracted version of spectral profile is also overlaid here (brown solid line).}
\label{figS6}
\end{figure}



\section{Angle dependent RIXS data}

In this section, we show incidence angle dependent RIXS results on the measured spin-orbital multiplet excitations and the elastic scattering at base temperature 20 K. In particular, we focus on the angular dependence of the spin-orbital excitations peak 1 and 2 in the antiferromagnetic state. In Supplementary Figure 5, the mode energies and spectral intensities of peak 1 and peak 2 are plotted in both $\pi$ and $\sigma$ incident light polarization. The spectral components are fitted by pseudo-Voigt functions as described in the main text. These measurements are carried out by varying the incident X-ray angle with respect to the sample surface while fixing the scattering angle 2$\theta$. With this procedure, we infer that both peak 1 and 2 exhibit flat energy dispersion as compared to their broad spectral width. Some finite energy differences in these assigned multiplet manifolds with varied angles are due to RIXS cross-section changes with scattering geometry, which depends on the incident light angle and polarization \cite{DeGroot2021}. Notably, here we express the changes by the correspondent incident angle for referring to such scenario. On the other hand, the spectral intensities for peak 1 and 2 are plotted similarly with angle and compared with the elastic scattering. The overall sinusoidal-like trend for peak 1, peak 2 and elastic line agree with the combination of saturation and self-absorption effects in previous reports on multiplet excitations in Co$^{3+}$ 3$d^6$ compounds \cite{Wang2020b}. Namely, the scattered photons can get re-absorbed differently with varied incident angles, imposing spectral distortions on top of the RIXS cross-section. The actual absorption coefficients are nearly constants in our experimental configuration. 

\begin{figure}[!htbp]
\centering
\includegraphics[width=\linewidth]{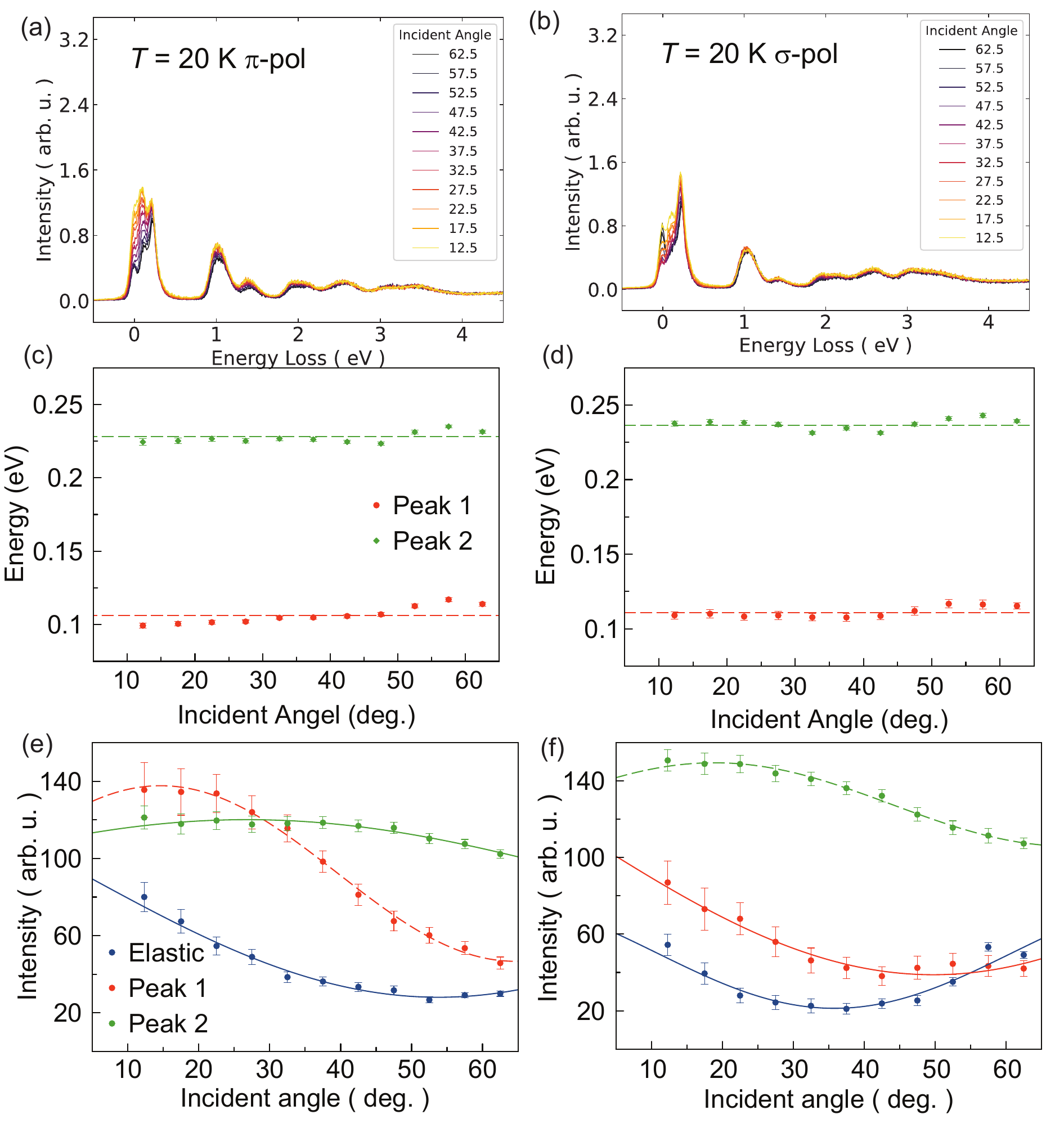}
\caption{Momentum dependent RIXS data for spin-orbital multiplet excitations peak 1 and 2. a-b: Displaying raw data for different incident angles. c-d, The fitted mode energies and e-f, spectral intensities are plotted as function of the incident angle of X-rays with respect to the sample surface.}
\label{figS5}
\end{figure}




\section{Additional charge-transfer multiplet theory calculations}
Here we provide additional details of the charge transfer multiplet (CTM) calculations. Particularly, we elaborate on the hypothesis of enhanced spin-singlet multiplet ground states in the paramagnetic phase, as well as the alternative physical picture for temperature dependence of multiplet excitations.




\subsection{Hypothesized singlet $^5$A$_1$ ground state in paramagnetic state}

In this section, we introduce a possible scenario to explain our observation for temperature evolution of enhanced zero-energy elastic peak when heating above the antiferromagnetic ordering temperature. In Figure 2 and 4 of main text, we find that the zero-energy elastic line exhibits increased intensity with a steep jump in vicinity to the antiferromagnetic transition $\sim$120 K. Here we postulate a mechanism based on a singlet ground state to explain this phenomenon. This has been observed in multiplet ground and excited state transitions of other magnetically-ordered systems, where the spin- and orbital-singlet components at zero-energy loss are enhanced with higher state symmetry in accordance to the magnetic (and lattice) structure in the high-temperature paramagnetic phase \cite{Miao2019a,Marino2023}.

Our tests with CTM theory are summarized in Supplementary Figure 6. Crucially, to reproduce such zero-energy singlet ground states, we conclude that an enhanced weight of multiplet excitations in $\sim$400-550 meV should be present. This has been hinted in a previous neutron study \cite{Rule2009} and indeed observed in our RIXS experiments, where some inelastic spectral contribution must be accounted in this energy regime (see fitting component about 350-400 meV in Supplementary Figure 4). Nevertheless, it only appears as residual weight that gets overwhelmed by the higher-energy tail from the predominant multiplet peak 2 ($\sim$220 meV; see Supplementary Figure 2) in the experimental RIXS response. On the other hand, the RIXS elastic signal involves multiple other contributions that cannot be solely and predominantly contributed by the multiplet excitations, e.g. resolution-limited excitations such as magnons/phonons, diffuse scattering out of surface roughness, etc. Consequently, a thorough assessment of the zero-energy (quasi-) elastic scattering as such would require further investigations, as all these factors could account for systematic and experimental uncertainties.

\begin{figure}[!htbp]
\centering
\includegraphics[width=\linewidth]{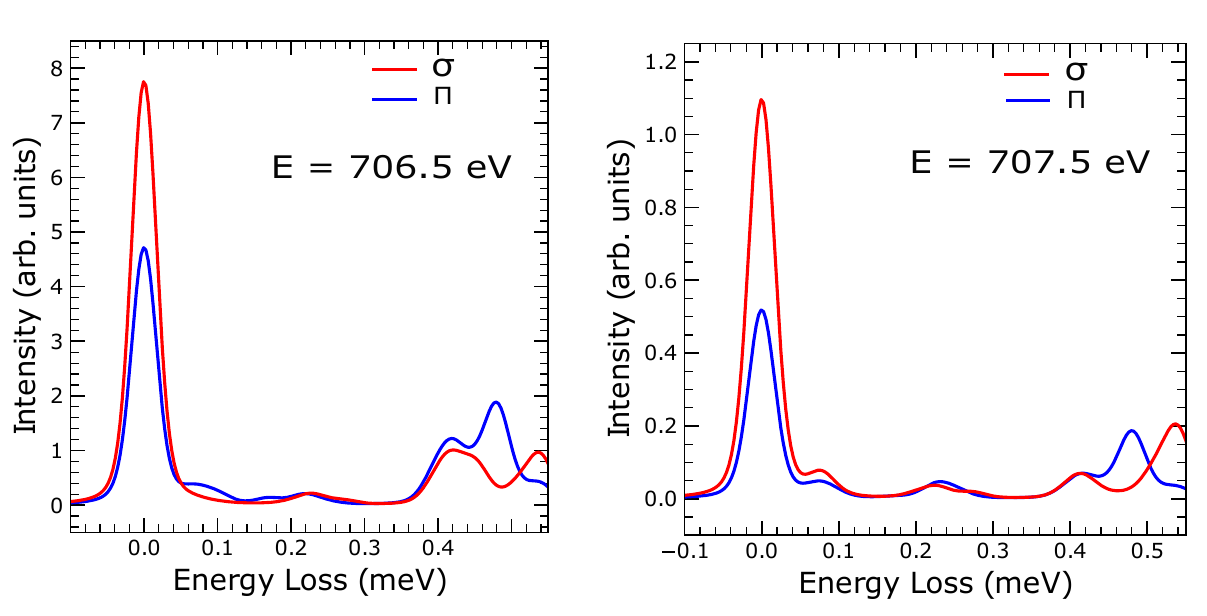}
\caption{Fe $2p3d$ RIXS calculations for FePS$_3$ for parameters leading to a singlet ground state ($^5$A$_1$) for incident energies E = 706.5 eV (left) and E = 707.5 eV (right). The polarization of the incident beam is calculated for $\sigma$ (blue) and $\pi$ (red). The distortion parameters used for the calculations are 10D$_q$ = 1 eV, D$_s$ = -60 meV and D$_t$ = 50 meV. Significant $dd$ excitations can be seen in the energy loss range from 400 to 550 meV, which merge with the higher-energy tail weight of the multiplet peak 2 ($\sim$220 meV) in our experiment.}
\label{figS1}
\end{figure}

\subsection{Alternative thermal development scenario for the spin-orbital multiplet excitations}
In this section, we show another possible picture for the temperature dependence of the measured multiplet excitations in Supplementary Figure 7. In the main text Figure 4, we compare the experimental RIXS response with respect to CTM calculations in which the magnetic exchange interactions are switched off in the paramagnetic phase. By this procedure, we reproduce the two-state crossover like spectral evolution of the RIXS response across the magnetic transition as a function of temperature. 

On the other hand, the other scenario of magneto-restriction is not sufficient to lead to correspondent changes in magnetic exchange couplings. For the case of constant exchange interactions below and above the antiferromagnetic phase transition, the spectral changes with temperature are mostly driven by thermal populations following Boltzmann statistics, as these optically-forbidden multiplet transitions can naturally couple to bosonic exciations (predominantly phonons here). This would give only a continuous trend displayed in Supplementary Figure 7 that is insensitive to the magnetic phase transitions, or other broadening factors if any.

\begin{figure}[!htbp]
\centering
\includegraphics[width=\linewidth]{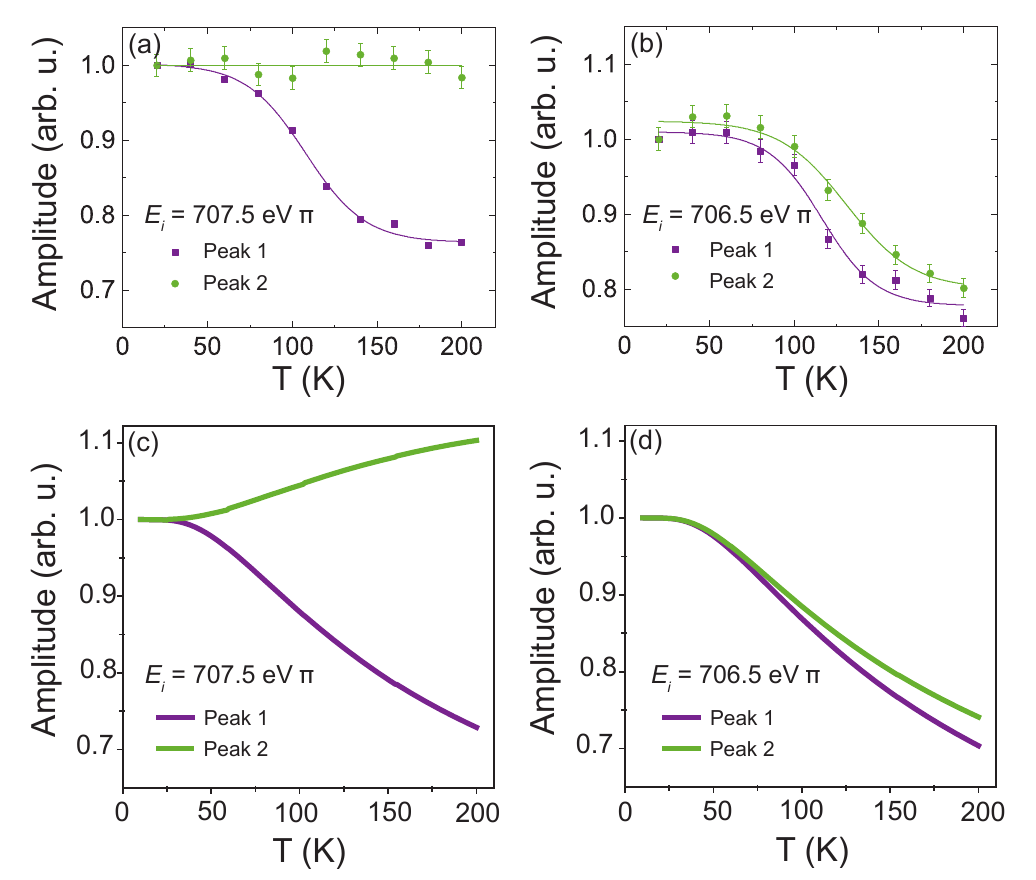}
\caption{Temperature dependence of the RIXS intensity of the spin-orbital excitations as we cross the magnetic transition temperature. (a) and (b) experimental results. (c) and (d) theoretical calculations where the exchange interaction is kept constant for all the temperatures.}
\label{figS2}
\end{figure}

\section{Characterization of mechanically exfoliated flake samples}
FePS$_{3}$ thin flake samples were meticulously prepared through 'scotch-tape' mechanical exfoliation of single crystalline materials. For the specific purpose of conducting X-ray measurements, isolated flake samples were meticulously crafted via van der Waals force transfer. The process entailed initially mechanically exfoliating the sample onto a silicon substrate, followed by the precise selection of flakes. Subsequently, these selected flakes were delicately transferred onto a stack comprising a polycarbonate film atop a polydimethylsiloxane (PDMS) stack. The polycarbonate film was later removed using a chloroform solution. These exfoliated crystals exhibited remarkable stability when exposed to ambient air conditions, with no discernible alterations observed under optical microscopy over a period of at least one week. To achieve the desired mask pattern, the areas enclosed by the mask were meticulously etched using a precision needle tip. The thickness of the monolayer is 0.78 nm \cite{Wang2016}, indicating that the flake is composed of 4-5 layers.

To measure the spatially separated flake samples, we employed an aperture horizontal slit at the exit slit position of the upstream beamline optics to further trim the X-ray beam focusing, achieving a footprint of $\sim$20$\times$4 $\mu$m$^{2}$ (horizontal$\times$vertical). For the 50 ML and 5 ML flake samples, the X-ray measurements were performed with a scattering angle 2$\theta$ = 130$^{\circ}$ for the experimental geometry. The scattering plane is fixed to the bc plane with the in-plane momentum transfer along the crystallographic [010] direction in the normal incidence configuration. $\sigma$ polarization was employed for the incident X-rays.

\begin{figure}[!htbp]
\centering
\includegraphics[width=\linewidth]{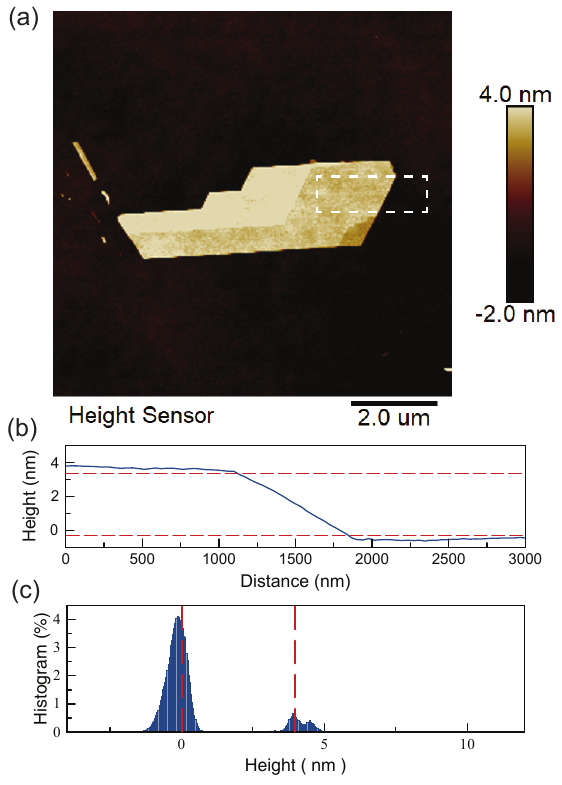}
\caption{(a) Atomic force microscopy image of exfoliated FePS$_{3}$ samples. The white dash box indicates the height measurement area. (b) The height difference in the direction across the edge. The red dotted line is the visual extension of the height step. (c) The histogram of the heights in the scanning area, the red dash lines are guided to eyes.}
\label{figS8}
\end{figure}

\bibliography{FePS3_supp}